\documentclass[final,1p,times]{elsarticle}

\usepackage{lineno,hyperref}
\modulolinenumbers[5]
\usepackage{multirow}
\usepackage{amsmath} 
\usepackage{amssymb} 
\usepackage{graphicx}
\usepackage{dcolumn}
\usepackage{bm}
\usepackage{adjustbox}
\usepackage[utf8]{inputenc}

\biboptions{sort&compress}

\newcommand{\jpsi}{J/\psi}
\newcommand{\dint}{{\rm d}}
\newcommand{\as}{\alpha_{\rm s}}
\newcommand{\aem}{\alpha_{\rm em}}
\newcommand{\sartre}{Sar\emph{t}re}
\newcommand{\xpom}{x_{\!I\!\! P}}

\begin{document}

\begin{frontmatter}

\title{Investigating saturation effects in ultraperipheral collisions at the LHC with the color dipole model}

\author[a]{Bharath Sambasivam}
\author[b]{Tobias Toll}
\ead{ttoll@bnl.gov}
\author[c]{Thomas Ullrich}

\address[a]{Department of Physics, Syracuse University, Syracuse, NY, USA}
\address[b]{Indian Institute of Technology Delhi, Hauz Khas, New Delhi, India}
\address[c]{Department of Physics, Brookhaven National Laboratory, Upton, NY, USA}

\begin{abstract}
We investigate saturation effects in $ep$ scattering as well as in ultraperipheral $p$A and AA collisions at small $x$ with four variants of the impact parameter dependent color dipole model: with and without gluon saturation and with and without a novel mechanism that suppresses  unphysical dipole radii above the confinement scale, a problem not addressed by most implementations. We show that $ep$ scattering at HERA can be very well described by any of the four variants. When going from $ep$ to $e$A scattering, saturation effects are expected to increase as $\sim$A$^{1/3}$. In lieu of an electron-ion collider, we confront the different versions of the dipole model with data recorded in ultraperipheral collisions at the LHC in order to estimate the sensitivity of the  data to gluon saturation in the target nuclei. We find that ultraperipheral PbPb collisions indicate strong saturation effects while $p$Pb collisions  turn out to not have any discriminating power to distinguish saturation from non-saturation scenarios.
\end{abstract}


\end{frontmatter}



\section{\label{sec:intro} Introduction}
Over the last two decades the Color Dipole Picture \cite{GolecBiernat:1998js, GolecBiernat:1999qd, Kowalski:2003hm, Kowalski:2006hc, Rezaeian:2012ji, Mantysaari:2018nng} has been developed to study high energy scattering in QCD. An attractive feature of the dipole approach to high-energy interactions is that it gives a clear interpretation of the physics at small values of Bjorken $x$.  Even in its simplest form, it turns out to be rather successful in describing the total \cite{GolecBiernat:1998js, Navelet:1995fa, Navelet:1996jx} and diffractive \cite{GolecBiernat:1999qd, Bialas:1997vt, Munier:1998nj} virtual photon-nucleon cross-sections. 
Its generalizations are now commonly used for the parametrization of data from HERA \cite{Luszczak:2013rxa, Luszczak:2016bxd, Mantysaari:2018nng}. In the scattering of leptons off hadrons at high energy, high-density phases of partons are created and non-linear effects become important. In this “saturation” regime, the basic properties of perturbative QCD, such as collinear factorization and linear evolution, break down. This unique form of strongly interacting matter is called a Color Glass Condensate (CGC) \cite{Iancu:2003xm}. The color dipole model has been originally created to study scattering in this regime.

From a phenomenological point of view, the key is to connect the experimental probes to the scattering of a dipole. In the case of lepton-hadron collisions this is obvious: a lepton undergoes hadronic interactions via a virtual photon and one can regard the interaction as the fluctuation of the virtual photon into a  quark-antiquark pair which then interacts. In hadron-hadron collisions however, there are no such virtual-photon probes. The exception are ultraperipheral collisions (UPC) where, at very large impact-parameters between the colliding hadrons, the long range electromagnetic force becomes dominant over short-range QCD.  The intensity of the electromagnetic field, and therefore the number of photons in the cloud surrounding the nucleus, is proportional to the square of the hadron's electric charge. Thus these types of interactions are highly favored when heavy ions collide. UPC are currently extensively measured at RHIC \cite{Adler:2002sc, Adams:2004rz, Abelev:2007nb, Adamczyk:2017vfu} and LHC 
\cite{Abelev:2012ba, Abbas:2013oua, TheALICE:2014dwa, Acharya:2018jua, Adam:2015gsa, Adam:2015sia, Khachatryan:2016qhq, Sirunyan:2018sav, LHCb:2018ofh}. 

Various attempts have been made to describe UPC data in the dipole model to test if saturation effects are present
\cite{Lappi:2013am, Santos:2014zna, Ducati:2016jdg, Luszczak:2017dwf, Mantysaari:2017dwh, Goncalves:2017wgg, Goncalves:2018blz, Kopp:2018xvu, Luszczak:2019vdc}. 
The parameters that provide the necessary non-perturbative input to these models are commonly determined through fits to high-precision structure functions measured at HERA in $ep$ collisions \cite{Kowalski:2006hc, Rezaeian:2012ji, Mantysaari:2018nng}.  So far, the comparison of color dipole models with UPC data has provided no clear evidence for or against saturation. Failure (or success) in describing UPC data with the dipole model does not imply the absence (or presence) of saturation but could also point to shortcoming of the model itself. Improvements, such as NLO calculation, are under way \cite{Boussarie:2016bkq, Beuf:2017bpd}. 
Another issue, one we address in this paper, is that the dipole model allows for large dipole radii beyond the confinement scale which are unphysical. We will show how this affects the comparison with data using the diffractive event generator \sartre~\cite{Toll:2013gda, Toll:2012mb}. To overcome the issue of unphysical large radii we implemented  a damping mechanism inspired from what was originally introduced in \cite{Flensburg:2008ag}. To do so, new dipole parameters had to be obtained by fitting the modified dipole model to HERA data.

In this paper we also consider two versions of the dipole model, one with saturation (bSat\footnote{In literature,  bSat (bNonSat) is also known under the name IPSat (IPNonSat).}) and one without (bNonSat). The latter is obtained by linearizing the dipole cross-section. This comparison allows us to test the actual sensitivity of UPC data to saturation effects.
\section{\label{sec:level2}Nonlinear effects in the Dipole Model}
In Deep Inelastic Scattering (DIS) electron-hadron scattering a virtual photon interacts electromagnetically with a parton in the hadron. At small parton momentum fractions $x$, this process can be seen as the virtual photon fluctuating into a quark anti-quark color dipole which subsequently interacts via one or more gluon exchanges with the proton. This process is described by impact parameter dependent dipole model \cite{Kowalski:2003hm, Kowalski:2006hc}. The DIS process has the following cross-section:
\begin{eqnarray}
	\sigma^{\gamma^* p}_{\rm L, T} (x, Q^2)= \sum_f \int \dint^2{\bf b}\dint^2{\bf r} \int_0^1\frac{\dint z}{4\pi}
	\left|\Psi_{\rm L,T}^f(r, z, Q^2)\right|^2\frac{\dint\sigma_{\rm dip}}{\dint^2\bf{b}}. \nonumber
\end{eqnarray}
This can be seen as a three part process where the virtual photon splits into a color dipole, which interacts with the hadron and then recombines. The splitting and recombination of the virtual photon is described by the transversely and longitudinally polarized wave functions:
\begin{eqnarray}
	\left|\Psi_{\rm T}^f(r, z, Q^2)\right|^2 &=& 
	\frac{2N_C}{\pi}\aem e_f^2\big([z^2+(1-z)^2]\epsilon^2K_1^2(\epsilon r)\nonumber\\ 
	&~&+~m_f^2 K_0^2(\epsilon r)\big) \nonumber\\
	\left|\Psi_{\rm L}^f(r, z, Q^2)\right|^2 &=&  
	\frac{8N_C}{\pi}\aem e_f^2Q^2 z^2 (1-z)^2K_0^2(\epsilon r)
	\label{eq:waves}
\end{eqnarray}
where $\epsilon^2 = z(1-z)Q^2 + m_f^2$. Here, $z$ is the quark's momentum fraction of the photon, $m_f$ the quark mass, $Q^2$ the photon virtuality, $x$ the gluon's momentum fraction of the hadron, and $g$ is the DGLAP longitudinal gluon density \cite{Gribov:1972ri, Gribov:1972rt, Altarelli:1977zs, Dokshitzer:1977sg}. The interaction between the dipole and the hadron is described by the dipole cross-section, which can be seen as the exchange of one gluon \cite{Kowalski:2003hm}:
\begin{eqnarray}
	\frac{\dint\sigma_{\rm dip}}{\dint^2{\bf b}} = r^2\frac{\pi^2}{N_C}\as (\mu^2)xg(x, \mu^2) T_p(b)
\label{eq:bNonSat}
\end{eqnarray}
There are three types of phenomenological modifications in the dipole cross-section which address non-linear effects. Firstly, the proton thickness $T_p(b)$ cannot be calculated from first principle. However, for exclusive diffraction, the impact parameter $b$ is an observable, and it can be experimentally accessed via its Fourier conjugate, the Mandelstam variable $t$. Investigations of the $t$ distributions in exclusive $\jpsi$ production have shown that the thickness is well described by a Gaussian $T_p(b)=\exp(-b^2/2B_p)/2\pi B_p$, with the parameter $B_p=4~$GeV$^{-2}$ \cite{Kowalski:2006hc, Rezaeian:2012ji, Mantysaari:2018nng} . The proton's thickness will not be the focus of this paper.

It is further expected that the dipole size $r$ is dampened for radii larger than the confinement scale. There is an inherent damping effect in the Bessel functions in the wave overlap, since at large $r$ these are suppressed as $\sim \exp(-r\epsilon)$, giving an effective size of the dipole of $r\sim1/\epsilon$. If $\epsilon$ is small, which can happen at small $Q^2$ and at $z\simeq 0, 1$, this does not give enough suppression of large dipoles. The authors in \cite{Kowalski:2003hm, Kowalski:2006hc}  solve this by introducing artificially large masses of the light quarks similar to pion or $\rho$-meson masses ensuring a sizable $\epsilon$ analogous with a meson enhancement. Later approaches \cite{Rezaeian:2012ji, Mantysaari:2018nng} have ignored this problem and allowed for any $\epsilon$, and thereby allowed for unnaturally large dipoles.  A novel approach is to add a Gaussian suppression to the dipole cross-section by hand, inspired by Flensburg {\it et al.} \cite{Flensburg:2008ag}: 
\begin{eqnarray}
	r_{\rm soft}(r_{\rm pert})= R_{\rm shrink}\sqrt{\ln\left(1+\frac{r_{\rm pert}^2}{R_{\rm shrink}^2}\right)}
\end{eqnarray}
where $r_{\rm soft}$ is the modified dipole size used in the dipole cross-section, $r_{\rm pert}$ is the dipole size given by the wave overlap. For small dipoles, $r_{\rm soft}\approx r_{\rm pert}$ as expected. Here $R_{\rm shrink}$ is a free parameter. This is the approach we will adopt in this paper.  It is worth noting that there are two effects of damping that have the opposite effect on the dipole cross-section. 
The shift $r_{\rm pert} \rightarrow r_{\rm soft}$, where $r_{\rm soft} <  r_{\rm pert}$, causes a trivial, $x$-independent suppression due to the $r^2$ factor in ${\dint\sigma_{\rm dip}}/{\dint^2{\bf b}}$.
However, the shift in $r$ also affects $xg(x,\mu^2)$ since $\mu^2=C/r^2+\mu_0^2$.  The increase in the scale, $\sim 1/r_{\rm soft}^2$, increases typically the gluon density thus increasing the cross-section. This  introduces also an $x$ dependence to the damping.

The third consideration is related to saturation, which also suppresses large dipoles as well as very small values of $x$. Saturation effects are important in the region of phase-space where the gluon wave-functions are overlapping, and large dipoles have an increased amplitude for interacting with multiple gluons and therefore pick up their correlations. Also, at small $x$, the gluon wave-functions becomes spatially larger and therefore overlap more. The saturated version of the dipole model may in principle be derived from the Color Glass Condensate effective theory for QCD \cite{Gelis:2010nm}, and is of the following form:
\begin{eqnarray}
	\frac{\dint\sigma_{\rm dip}}{\dint^2{\bf b}} =2\left[ 1-\exp\left(-r^2\frac{\pi^2}{2N_C}\as (\mu^2)xg(x, \mu^2) T_p(b)\right)\right]
\label{eq:bSat}
\end{eqnarray}
At small $r$, this expression becomes equal to Eq.~\eqref{eq:bNonSat}. At first glance, this too seems to give a Gaussian suppression for large dipoles. However, the $r$-dependence in the scale of the DGLAP gluon density makes the suppression more involved. In Eq.~\eqref{eq:bSat} large gluon densities are also suppressed as expected. Eq.\eqref{eq:bSat} is referred to as the bSat model, while Eq.~\eqref{eq:bNonSat} is referred to as the bNonSat model.

In order to disentangle large dipole effects we believe that we need to have high precision measurements of the gluon density in the nucleus. When going from $ep$ to $eA$ collisions, the confinement scale is expected to remain unchanged while the saturation scale gets an ''oomph`` of $\sim A^{1/3}$. Thus, by investigating the $A$ dependence of the cross-section one should be able to disentangle these non-linear effects.  This will be one of the main foci of the future Electron-Ion Collider (EIC) \cite{Accardi:2012qut} which will be constructed in the US in the next decade. In lieu of an EIC,  ultraperipheral collisions measured at RHIC and LHC offer the possibility to study similar processes in photoproduction, where $Q^2\approx 0$. In this paper we will attempt to disentangle the large dipole effects using combined $e^\pm p$ DIS data from HERA I+II as well as $p$Pb and PbPb UPC data for exclusive vector meson production at the LHC.  For this purpose we have implemented the calculations in the event generator \sartre~ \cite{Toll:2012mb, Toll:2013gda}.  
\section{\label{sec:level1}Fits to inclusive DIS at HERA}
The proton structure functions $F_2(x, Q^2)$ and $F_L(x, Q^2)$ can be written in terms of the total photon-proton cross-sections as:
\begin{eqnarray}
	F_2(x, Q^2)&=&\frac{Q^2}{4\pi\aem}(\sigma_{\rm T}^{\gamma^*p} + \sigma_{\rm L}^{\gamma^*p} ) \nonumber \\
	F_L(x, Q^2)&=&\frac{Q^2}{4\pi\aem}\sigma_{\rm L}^{\gamma^*p}  \nonumber 
\end{eqnarray}
Following \cite{Mantysaari:2018nng}, we take the DGLAP gluon density at a starting scale $\mu_0^2=1.1~$GeV$^2$ to be
	$xg(x, \mu_0^2)=A_gx^{-\lambda_g}(1-x)^6 \mathrm{,}$
using variable flavour scheme when evaluating the strong coupling $\as$ and solving the DGLAP evolution for the gluon density. The strong coupling satisfies $\as(\mu^2=M_Z^2)=0.1183$. For heavy quarks, the Bjorken $x$ is replaced by 
$x_f=x\left(1+\frac{4m_f^2}{Q^2}\right)$.
We fit to the combined reduced cross section data from HERA I and HERA II \cite{Abramowicz:2015mha, Aaron:2009aa} and the combined reduced cross section for charm data  \cite{Abramowicz:1900rp, H1:2018flt} from the H1 and ZEUS experiments. The reduced cross section is given by:
\begin{eqnarray}
\sigma_r(x,y,Q^2)=F_2(x, Q^2)-\frac{y^2}{1+(1-y)^2}F_L(x, Q^2)
\end{eqnarray}
We include data in the range $x<0.01$, or $x_f<0.01$, and $1.5\leq Q^2\leq 50~$GeV$^2$. This gives us 409 data points for $\sigma_r$ and 34 points for $\sigma_r^{c\bar c}$. The results from the fits including confinement are presented in Table \ref{tab:fit} as well as the result of the previous fit \cite{Mantysaari:2018nng} for comparison. When selecting the data points included in the fit, we used $m_c=1.321$ GeV, which  differs slightly from the choice in \cite{Mantysaari:2018nng}. This causes a negligible difference in the number of data points used for our fit.

\begin{table}
\begin{center}
  \begin{adjustbox}{width=\linewidth}
	\begin{tabular}{	|c|c|c|c|c|c|c|c|c|}
	\hline 
Model                                               & $\chi^2/$Ndf & N & $m_l$ (GeV)    & $m_c$  (GeV) & $C$    & $A_g$  & $\lambda_g$ & $R_{\rm shrink}$ (fm)\\\hline\hline 
bNonSat (damped)                           & {\bf 1.108}   & 409+34 & 0.05116 & 1.3446 & 1.7076 & 2.3938 & 0.06581         & 0.9025\\\hline
bSat (damped)                                  & {\bf 1.270}   & 409+34 & 0.004     & 1.4280 & 1.9724 & 2.1945 & 0.09593         & 1.1889 \\\hline
bNonSat \cite{Mantysaari:2018nng} & {\bf 1.317}   & 410+33 & 0.1497   & 1.3180 & 3.5445 & 2.8460 & 0.008336 & \\\hline
bSat \cite{Mantysaari:2018nng}       & {\bf 1.290}    & 410+33 & 0.03       & 1.3210 & 1.8178 & 2.0670 & 0.09575 & \\\hline
	\end{tabular}
	\end{adjustbox}
\caption{The resulting parameters from fitting to HERA I and HERA II data  \cite{Abramowicz:2015mha, Aaron:2009aa, Abramowicz:1900rp, H1:2018flt}. Here $B_p=4$ GeV$^{-2}$, $\mu_0^2=1.1$ GeV$^2$, $m_b=4.75$ GeV, and $m_t=175$ GeV. X+Y points means
X points for inclusive and Y points for charmed reduced cross section. The table also contains the results from the previous fit \cite{Mantysaari:2018nng} for reference. 
For inclusive DIS, bNonSat with damping has  $\chi^2/$Ndf=1.073, and bSat with damping has $\chi^2/$Ndf=1.262.}
\label{tab:fit}
\end{center}
\end{table}

We observe that the bNonSat model with damping fits the HERA data slightly better than the bSat model, although all models describe the data reasonably well. We also see that adding damping effects to the bSat model does not significantly alter the resulting fit quality. For fits with the explicit damping model, the light quark masses are substantially smaller, as expected. Also, for bNonSat, adding damping allows for a slower growth of the gluon density for small $x$, as seen by the values of $\lambda_g$. Our fit results indicate that there are no definitive hints for saturation effects in the HERA data, since the data can be equally well described using a Gaussian suppression of large dipoles. Since the fit procedure is identical to that used in \cite{Mantysaari:2018nng}, we refer to this reference for an in-depth discussion of the sensitivity of the result on assumed quantities such as proton profile width $B_p$, bottom quark mass $m_b$, starting scale for DGLAP evolution $\mu_0$, and the choice of wave function. 

\begin{figure}
	\centering
	\includegraphics[width=\linewidth]{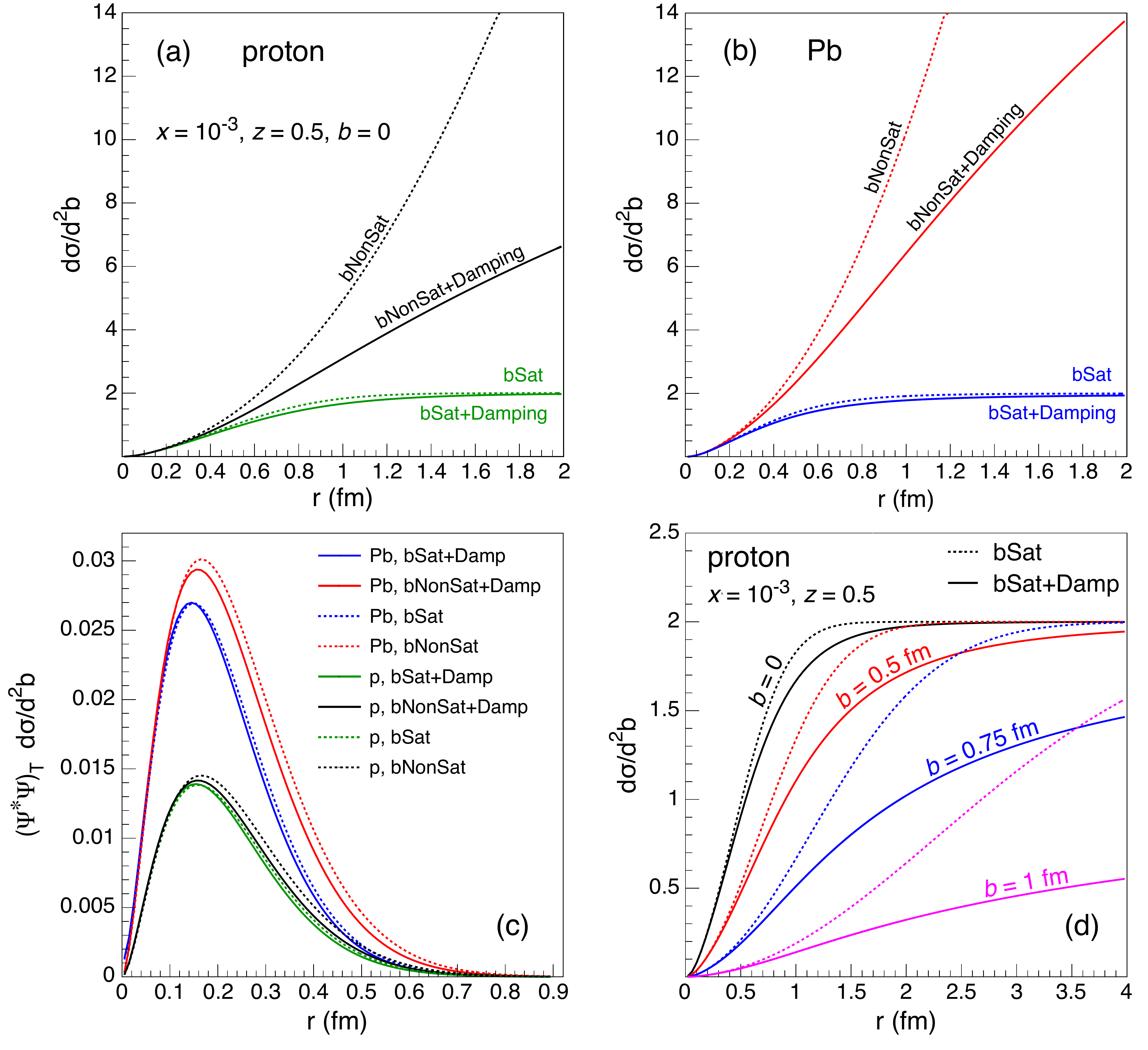}
	\caption{\label{fig:dipoles_p}
		A comparison of the resulting dipole models from table \ref{tab:fit}. (a) The proton dipole with and without damping and saturation, (b) the lead dipole. (c) The dipoles multiplied by the $\jpsi$ wave overlap for $Q^2=0$, (d) the proton dipole cross section at different impact parameters with and without damping. For all figures $x=10^{-3}$, $z=0.5$, in (a)-(c) $b=0$.}
\end{figure}

The resulting dipole cross-sections are shown in Fig.~\ref{fig:dipoles_p} (a) and (b) for proton and lead respectively, illustrating how the damping suppress large dipoles in the bNonSat model. We also note that adding damping to the bSat model increases the dipole cross-section slightly at smaller $r$. To anticipate the effects of damping and saturation in exclusive $\jpsi$ production in UPC, we also show in (c) the dipole cross-sections multiplied by the wave-function overlap between the incoming virtual photon and the produced $\jpsi$ meson, which by itself is also suppressing dipole radii larger than the typical size of the vector meson $r\sim 0.06$~fm. We are using the ``Boosted Gaussian" wave overlap parametrization from \cite{Kowalski:2006hc}. We note that for lead, both saturation and damping have significant impact, while for the proton the damping mechanism completely takes away any saturation effects. In Fig.~\ref{fig:dipoles_p} (d) we show how damping is affecting the saturated dipole cross-section for protons at different impact parameters, and we see that the damping effect becomes stronger at larger impact parameters. 
\section{\label{sec:level3}Exclusive Vector Mesons in UPC with \sartre}
In ultraperipheral collisions, the interacting hadrons are so far apart that the long range electromagnetic force dominates over the short range strong force. These interactions are therefore very similar to DIS with exchange of pseudo-real photons with $Q^2\approx 0$. For exclusive diffractive production of vector mesons in photoproduction the differential cross-section can be written as:
\begin{eqnarray}
\label{eq:cstot}
	\frac{\dint^2\sigma_{\rm total}}{\dint y\dint t}(\xpom, t)=
	(1+\beta^2)R_g^2 E_\gamma\frac{\dint n^\gamma}{\dint E_\gamma}
	\frac{1}{16\pi}|\mathcal{A}_{\rm T, L}(\xpom, t)|^2
\end{eqnarray}
where $y=\ln (2E_\gamma/\sqrt{m_V^2+p_{V\perp}^2})$, giving $\dint E_\gamma/\dint y=E_\gamma$ and $\xpom$ is the momentum fraction of the exchanged gluons with respect to the probed hadron.
Here $n^\gamma$ is the flux of photons interacting with the proton. In the derivation of the amplitude only the imaginary part is taken into account.  The correction for the missing real part  is given by $\beta=\tan(\lambda\cdot\pi/2)$, where $\lambda=-\partial\ln\mathcal{A}_{\rm T,L}/\partial\ln \xpom$ \cite{Kowalski:2006hc} . The  above cross-section treats the diffractive exchange as one or multiple two-gluon exchanges where the two gluons shield each other's colors, ensuring that the interaction does not change the proton's quantum numbers. However, to account for the possibility of the two gluons  having different momentum fractions $\xpom$, the amplitude is multiplied by a skewedness correction $R_g$ given by:
\begin{eqnarray}
	 R_g(\lambda_{\rm skew})=\frac{2^{2\lambda_{\rm skew}+3}}{\sqrt{\pi}}\frac{\Gamma(\lambda_{\rm skew}+5/2)}
	 {\Gamma(\lambda_{\rm skew} +4)},
\end{eqnarray}	  
where $\lambda_{\rm skew}=-\frac{\partial\ln(xg(x, \mu^2))}{\partial\ln x}$. As the skewedness corrections are formally needed only in the case of linear DGLAP gluon densities, and the dipole model modifies this density significantly, it is not theoretically clear whether or not it should be applied in this case. However, it has  historically been required  to describe $\jpsi$ production in $ep$ collision at HERA. In the following we will apply the skewedness correction to the proton dipole, where it is better motivated, but omit it for the ion dipoles. Both corrections become smaller with decreasing $x$. In the range relevant for this work, the real part correction is in the range $1+\beta^2=[1.17,1.53]$ while for the skewedness correction $R_g=[1.5,2.0]$.  In the case of $\gamma^*$A scattering, the bSat and bNonSat dipole cross-sections become:
\begin{eqnarray}
	&\frac{\dint\sigma_{\rm dip}^{A,{\rm bSat}}}{\dint^2{\bf b}}=
	2\left( 1-\exp(-r^2\frac{\pi^2}{2N_C}\as (\mu^2)xg\sum_{i=1}^AT_p(|{\bf b}-{\bf b_i}|))\right) \nonumber \\
	 &\frac{\dint\sigma_{\rm dip}^{A,{\rm bNonSat}}}{\dint^2{\bf b}}=
	r^2\frac{\pi^2}{N_C}\as (\mu^2)xg\sum_{i=1}^AT_p(|{\bf b}-{\bf b_i}|)
	\label{eq:dipoles_A}
\end{eqnarray}
where ${\bf b}_i$ are the positions of the nucleons in transverse space for a given nucleon configuration. 
The total cross-section is then given by:
\begin{eqnarray}
	\frac{\dint^2\sigma_{\rm total}}{\dint y\dint t}(\xpom, t)=
	(1+\beta^2) E_\gamma\frac{\dint n^\gamma}{\dint E_\gamma}
	\frac{1}{16\pi}\left<|\mathcal{A}_{\rm T, L}|^2\right>
\end{eqnarray}
where the average is taken over initial state nucleon configurations. The coherent part of the cross-section, in which the struck nucleus stays intact after the interaction, is given by:
\begin{eqnarray}
	\frac{\dint^2\sigma_{\rm coh}}{\dint y\dint t}(\xpom, t)=
	(1+\beta^2) E_\gamma\frac{\dint n^\gamma}{\dint E_\gamma}
	\frac{1}{16\pi}\left|\left<\mathcal{A}_{\rm T, L}\right>\right|^2
\end{eqnarray}
The incoherent case, where the struck nucleus becomes excited in the interaction and subsequently de-excites by emitting a photon, one or many nucleons, or fragments, is given by the difference between the total and coherent cross-sections. 
In \sartre~we generate 500 nuclear configurations and average the amplitude over these. This has been shown to be sufficient for the cross-sections to converge\cite{Toll:2012mb}. For the photon flux we follow the model used in the STARLIGHT generator \cite{Klein:2016yzr}. We have implemented these processes in the \sartre~event generator which give an exclusive final state with exact four-momenta of all incoming, intermediate and final state particles. 

\begin{figure}
	\includegraphics[width=\linewidth]{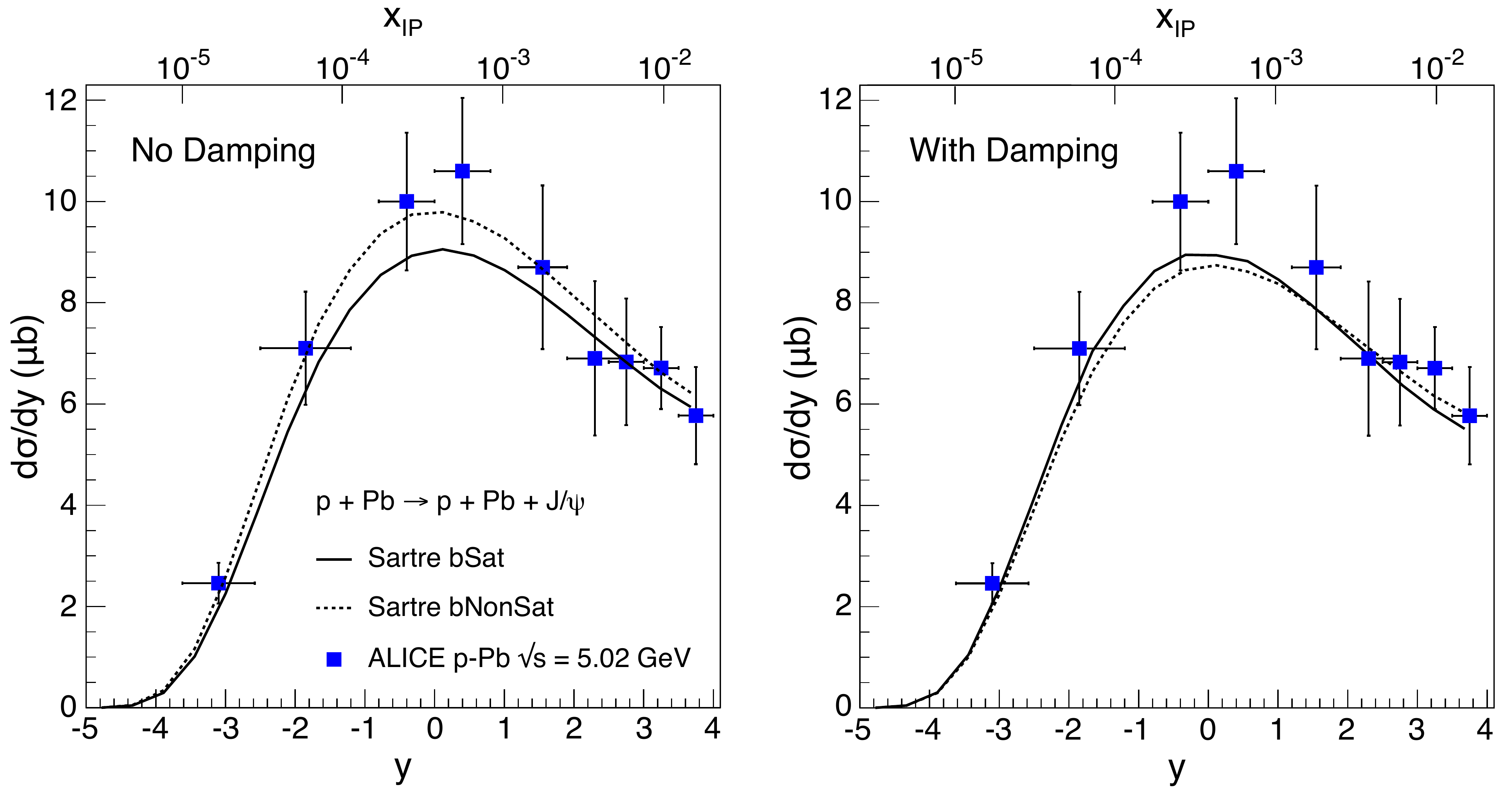}
	\caption{\label{fig:pPb}
	A comparison between \sartre~ and measured exclusive $\jpsi$ data from ALICE \cite{TheALICE:2014dwa, Acharya:2018jua} in $p$Pb UPC at $\sqrt{s}=5.02$~TeV. 
	We show the results using dipole models without (with) damping of large dipole radii on the left (right) hand side.}
\end{figure}

In Fig.~\ref{fig:pPb} we show cross-sections from two measurements in $p$Pb collisions by the ALICE collaboration \cite{TheALICE:2014dwa, Acharya:2018jua} with $\sqrt{s}=5.02$~TeV and compare them with results from the \sartre\ event generator. These measurements cover a rapidity range of $|y|<4$ which corresponds to $10^{-5} \leq \xpom \leq 10^{-2}$. In the left panel we show the \sartre\ results without damping, and in the right panel with damping, both in the same kinematic range as the data and integrated over $t$. We see that both versions of bSat are able to describe the data well. The bNonSat model without damping (left panel) deviates slightly more from bSat at all rapidities in the range, while the damped version of bNonSat (right panel) does not differ significantly from bSat. However, any differences are within the uncertainty range of the data.

\begin{figure}
	\centering
	\includegraphics[width=\linewidth]{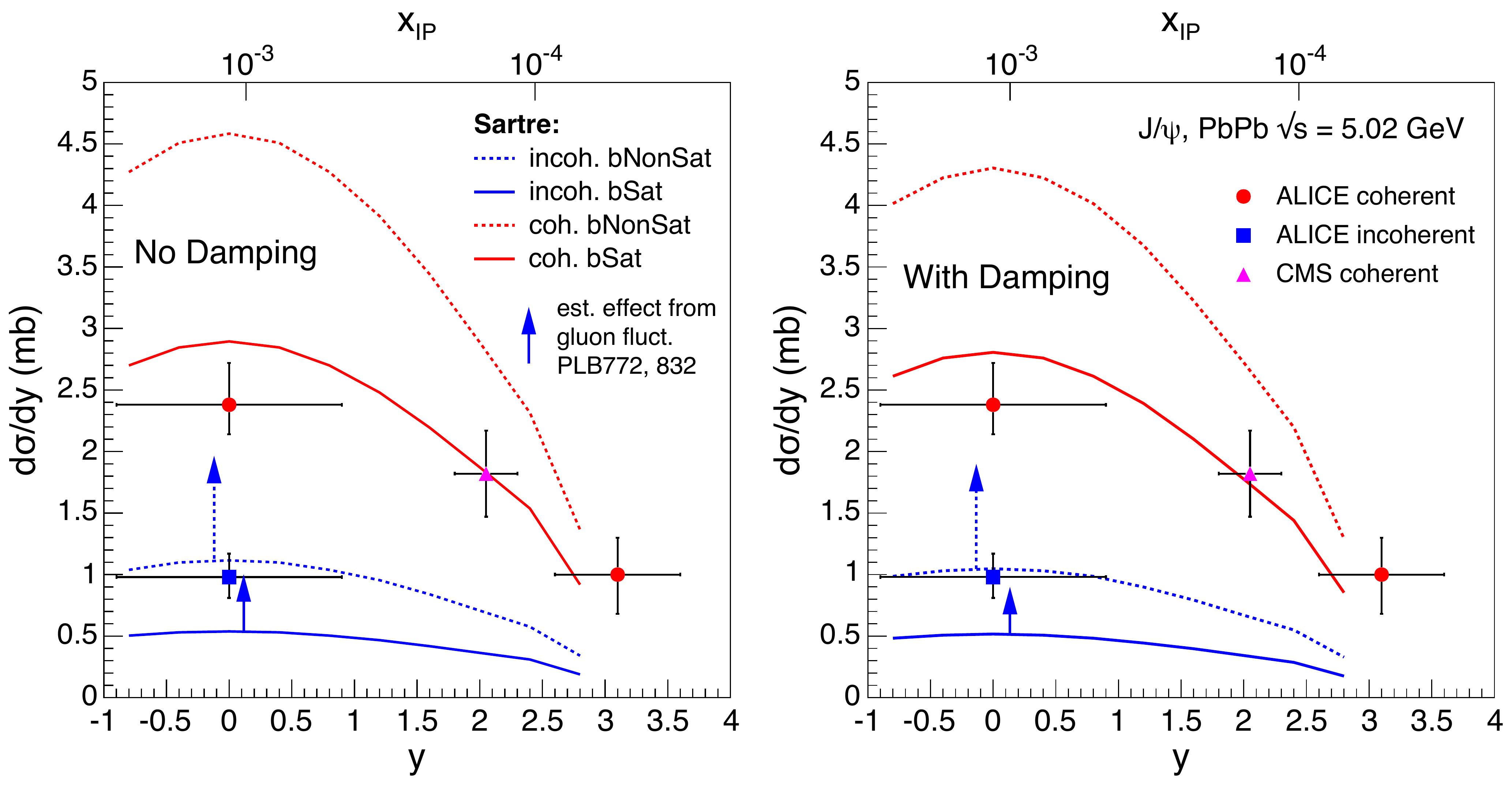}
	\caption{\label{fig:PbPb}
	A comparison between \sartre~and measured exclusive $\jpsi$ data from ALICE \cite{SCAPPARONE:2013isa} and CMS \cite{Khachatryan:2016qhq}
	of PbPb UPC at the LHC at $\sqrt{s}=2.76$~TeV. 
	The left (right) side show the results using dipole models without (with) damping. The arrows on the incoherent cross sections indicate where the curves would be if we included subnucleonic gluon fluctuations in the calculations, which according to \cite{Mantysaari:2017dwh} contributes a factor 1.8.}
\end{figure}

In Fig.~\ref{fig:PbPb} we show the results from events generated for ultraperipheral PbPb collisions at $\sqrt{s}=2.76$~TeV, and compare them with measurements from ALICE \cite{SCAPPARONE:2013isa} and CMS \cite{Khachatryan:2016qhq}. These measurements cover a rapidity range of $|y|<3.5$ which corresponds to $3\cdot 10^{-5} \leq \xpom \leq 3\cdot 10^{-2}$. Data and \sartre\ are integrated over $t$. For symmetrical beams, it is not experimentally possible to distinguish if the photon comes from beam 1 or beam 2, and the total cross-section is:
\begin{eqnarray}
	\frac{{\rm d}\sigma}{{\rm d}y}(y)=\frac{{\rm d}\sigma_1}{{\rm d}y}(y) + \frac{{\rm d}\sigma_2}{{\rm d}y}(-y)
\end{eqnarray}
Therefore, away from $y=0$ ($\xpom=10^{-3}$) the cross-section is a linear combination of larger and smaller $\xpom$. The dipole model is only applicable for $\xpom \lesssim 0.01$, which we relax a little since in this case it also imposes a lower bound on $\xpom$. \sartre~is therefore only able to access $|y|<2.9$ at this beam energy configuration. Both bSat versions can describe the coherent data well.  \sartre\ appears to favor a non-saturated model for the incoherent data as the bSat description lies below the data. However, this is to be expected, since the incoherent cross-section is proportional to gluon fluctuations, and \sartre~only includes fluctuations of the initial nucleon configuration in its model. We expect that this description will change once we have also included subnucleonic fluctuations and saturation scale fluctuations in our model following the prescription by M\"antysaari and Schenke \cite{Mantysaari:2016ykx, Mantysaari:2016jaz, Mantysaari:2017dwh}. In \cite{Mantysaari:2017dwh} the authors show that subnucleonic fluctuations increases the incoherent cross section in this beam configuration by a factor of 1.8. This is indicated by arrows in Fig.~\ref{fig:PbPb}.
We see that even with damping there is a large difference between bSat and bNonSat, at $y=0$ which corresponds to $\xpom=10^{-3}$, as anticipated from Fig.~\ref{fig:dipoles_p} (c). This difference is even larger in the incoherent cross-section. 

\begin{figure}
\begin{center}
	\centering
	\includegraphics[width=\linewidth]{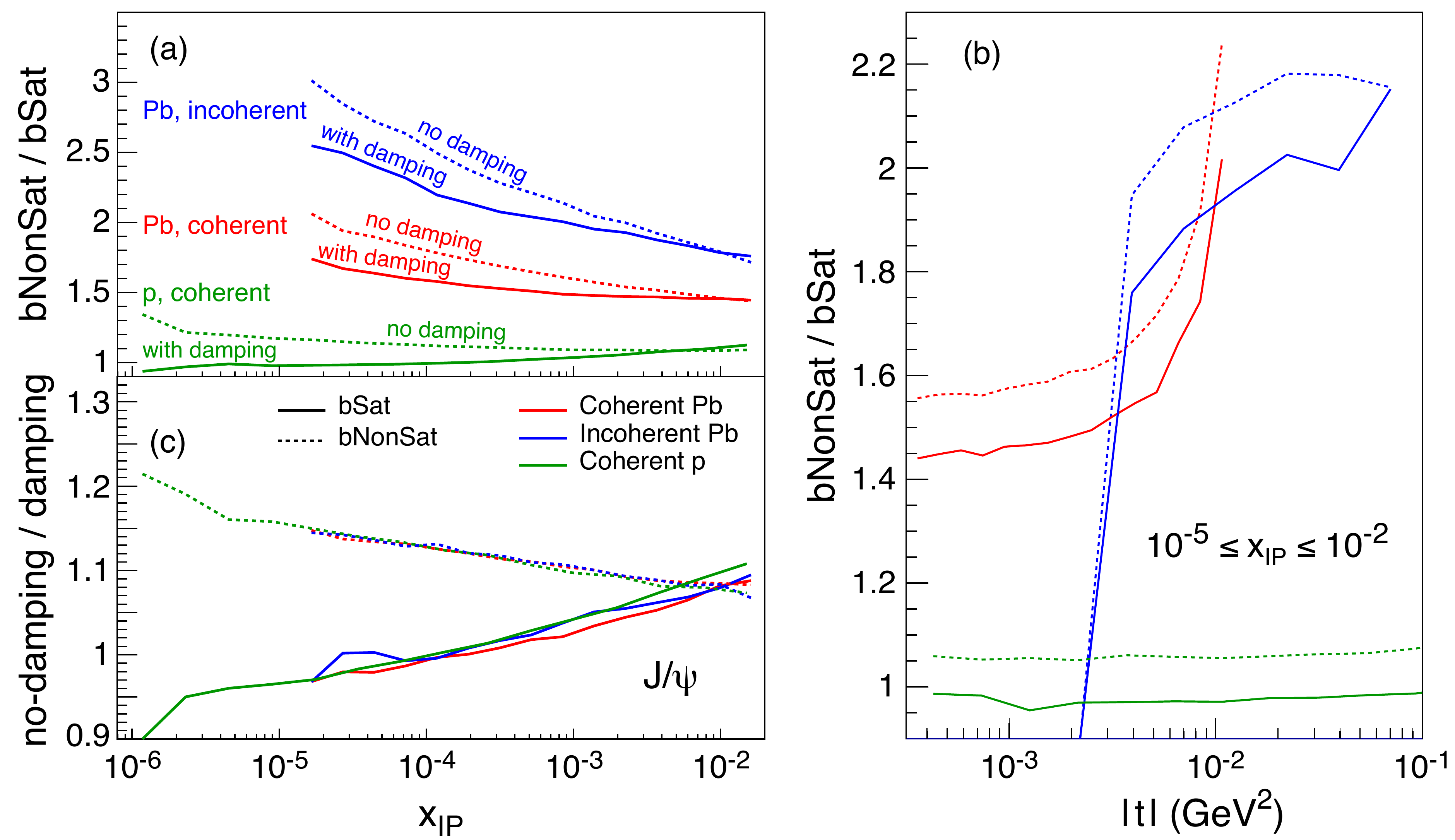}
	\caption{\label{fig:xpom}
	(a) Ratio of non-saturated to saturated cross-sections in exclusive $\jpsi$ production with and without damping of large dipole radii as a function of $\xpom$. (b) Same ratio but as a function  of $t$.  (c) Ratio of cross-section without dipole-radius damping over that with damping as a function of $\xpom$.}
\end{center}
\end{figure}

In Fig.~\ref{fig:xpom} (a) we show the non-saturated to saturated cross-section ratios as a function of $\xpom$  in exclusive $\jpsi$ production for incoherent $e$Pb and coherent $ep$ and $e$Pb collisions. We see that the largest saturation effects occur in the incoherent $e$Pb cross-sections, while for $ep$ there is little sign of saturation, especially in damping mode. In Fig.~\ref{fig:xpom} (b) we  depict the same ratio as a function of $t$ for  $10^{-5} \leq \xpom \leq 10^{-2}$. Note that the coherent cross-section in the bSat models decreases faster as a function of $|t|$ than in the bNonSat models. 
We further note that at larger $|t|$, where the incoherent part of the cross section dominates, the ratio becomes large.  For the proton, the ratio remains near unity for the entire $t$-spectrum. The large difference between bNonSat and bSat in Fig.~\ref{fig:PbPb} comes from the integral over all $t$. 

Fig.~\ref{fig:xpom} (c) illustrates the ratio of cross-section without dipole-radius damping over that with damping as a function of $\xpom$. We see that the $\xpom$ dependence of the damping is small and verify that it is independent of the nuclear species. The damping effects are identical for coherent and incoherent scattering. Naively the damping is not expected to have any dependence on $\xpom$. However, the absence of damping in the dipole cross section is compensated for by a smaller value of $\lambda_g$ in the fit, causing a slower growth in the gluon density at small $\xpom$, which can be seen in Table \ref{tab:fit}. There is also a direct dependence on the dipole radius, and therefore on the damping, in the factorization and renormalization scales $\mu^2=C/r^2+\mu_0^2$, which gives a larger effect in the DGLAP evolved gluon density at small $\xpom$ and large $\mu^2$. This is further enhanced by the $\jpsi$ wave function overlap. What is seen in Fig.~\ref{fig:xpom} is the combination of these effects. 

\begin{figure}
\begin{center}
	\centering
	\includegraphics[width=0.7\linewidth]{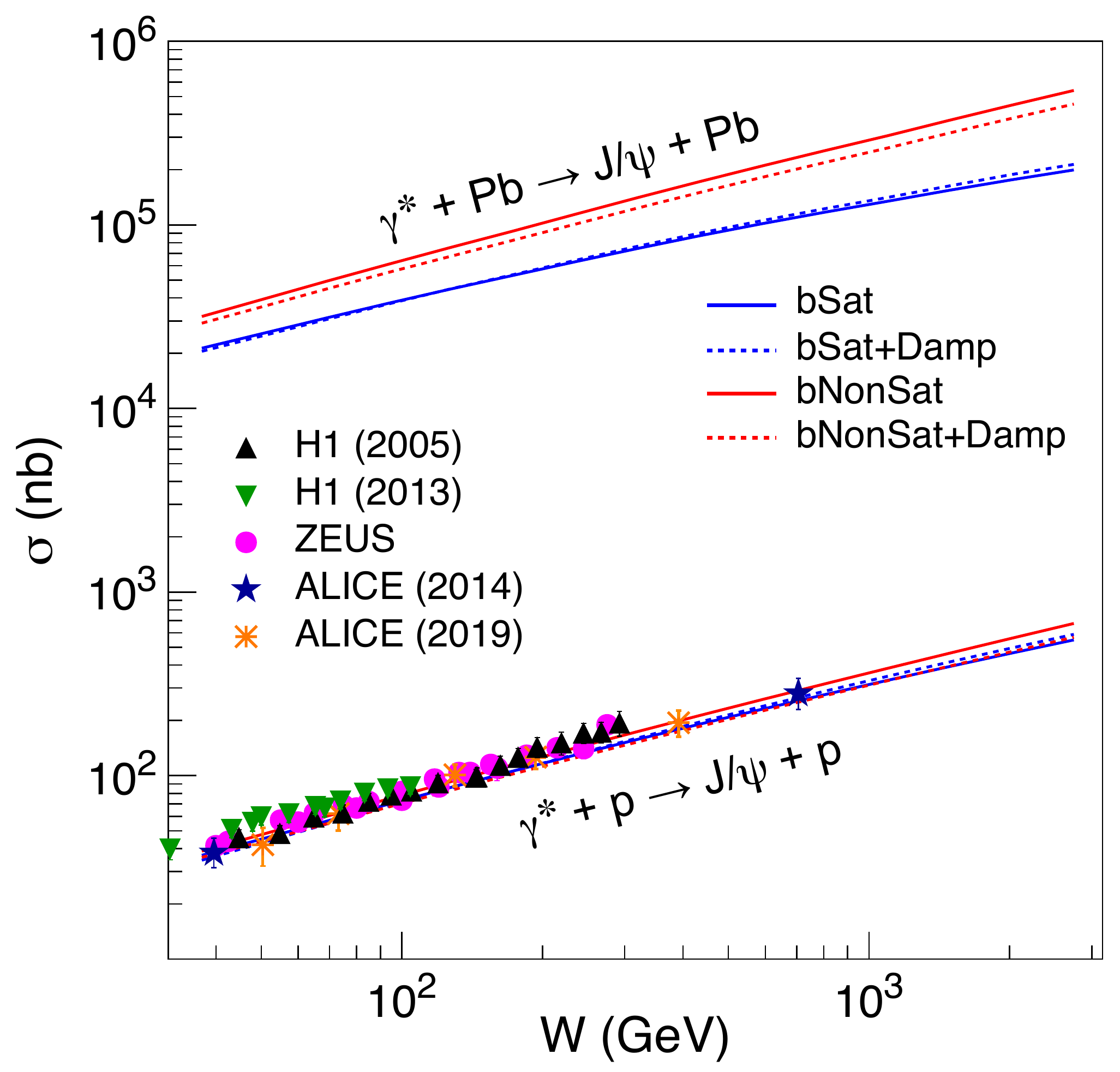}
	\caption{\label{fig:W}
	A comparison of the $W$ dependence of the different models for  $\gamma^{*}+p\rightarrow \jpsi +p$ and $\gamma^{*}+Pb\rightarrow \jpsi +Pb$. The proton case is compared to measurements from H1 \cite{Aktas:2005xu, Alexa:2013xxa}, ZEUS\cite{Chekanov:2002xi} and ALICE\cite{TheALICE:2014dwa, Acharya:2018jua} collaborations. All cross-sections are integrated over $t$.}
\end{center}
\end{figure}

In Fig.~\ref{fig:W} we show a comparison  of the  $W$-dependence of the total cross-section for exclusive $\jpsi$ photoproduction, $\gamma^{*}+p\rightarrow \jpsi +p$, between our four model variants and measurements from the H1 \cite{Aktas:2005xu, Alexa:2013xxa}, ZEUS\cite{Chekanov:2002xi} and ALICE\cite{TheALICE:2014dwa, Acharya:2018jua} collaborations. In addition we 
depict our predictions for $\gamma^{*}+p\rightarrow \jpsi + \mathrm{Pb}$, which will be tested at a future EIC. Again, we see that the difference between the models in $\gamma^* p$ interactions is not large enough for the data to distinguish between the saturation and non-saturation scenario, while in $\gamma^*$Pb  there is a pronounced separation
between the two. 

\section{\label{sec:conclusions}Conclusions}
We have introduced a damping for dipole radii larger than the confinement scale in the bSat and bNonSat models to improve the color dipole model. We show that this new model with newly derived parameters can describe HERA data well. The dipole model is implemented in the \sartre~event generator.  Our studies demonstrate that the damping improves the description of the non-saturated dipole model for ultraperipheral $p$Pb collisions at the LHC. However, the precision of the current data does not allow to discriminate between saturation and  non-saturation scenarios for $p$Pb collisions. 

The improved model describes ultraperipheral PbPb collisions quite well. 
Saturation effects in coherent PbPb collisions appear to be significant for all rapidities. A comparison of data from incoherent interactions will need to be further improved to include gluon fluctuations in the nucleons.

\section*{Acknowledgement}
We thank H. M\"antysaari and T. Lappi for fruitful discussions.
The work of T.U. is supported by the U.S. Department of Energy under Award DE-SC0012704. The work of T.T. was supported in part by Jefferson Lab LDRD Funding, LD1706 and LD1804, under US DoE contract DE-AC05-06OR23177. B.S. and T.T. thank the Physics Department at Shiv Nadar University.

\bibliography{bibliography}

\end{document}